\documentclass[a4paper,fleqn]{cas-sc}

\usepackage[numbers]{natbib}

\def\tsc#1{\csdef{#1}{\textsc{\lowercase{#1}}\xspace}}
\tsc{WGM}
\tsc{QE}
\tsc{EP}
\tsc{PMS}
\tsc{BEC}
\tsc{DE}

\newcommand{\toolname}{\texttt{HiP}\xspace}
\newcommand{\largestcluster}{$C_0$}

\begin{document}
\let\WriteBookmarks\relax
\def\floatpagepagefraction{1}
\def\textpagefraction{.001}

\shorttitle{Hesperus is Phosphorus}

\shortauthors{Roa, Suarez, and Tapiador}

\title [mode = title]{Hesperus is Phosphorus: Mapping Threat Actor Naming Taxonomies at Scale}                      
\tnotemark[1]

\tnotetext[1]{
This research was supported by MICIU/AEI/10.13039/501100011033 under Grant
No. PID2022-140126OB-I00 (CYCAD) and INCIBE under Grant APAMCiber. The opinions, findings, and conclusions or recommendations expressed are those of the authors and do not necessarily reflect those of any of the funding agencies.
}

\author{Gonzalo Roa}
\ead{100534780@alumnos.uc3m.es}
\credit{Data curation, Methodology, Software, Analysis, Writing, Visualization}

\author{Manuel Suarez-Roman}[orcid=0009-0008-2569-6178]
\ead{mansuare@pa.uc3m.es}
\credit{Data curation, Methodology, Software, Analysis, Writing, Visualization}

\author{Juan Tapiador}[orcid=0000-0002-4573-3967]
\ead{jestevez@inf.uc3m.es}
\credit{Conceptualization, Methodology, Software, Analysis, Writing, Visualization, Supervision, Project administration, Funding acquisition}
\cormark[1]

\affiliation[]{organization={Universidad Carlos III de Madrid},
    addressline={Avenida de la Universidad, 30}, 
    city={Leganés},
    postcode={28911}, 
    country={Spain}}

\cortext[cor1]{Corresponding author}

\begin{abstract}
This paper studies the problem of Threat Actor (TA) naming convention inconsistency across leading Cyber Threat Intelligence (CTI) vendors. The current decentralized and proprietary nomenclature creates confusion and significant obstacles for researchers, including difficulties in integrating and correlating disparate CTI reports and TA profiles. This paper introduces \texttt{HiP} (Hesperus is Phosphorus, a reference to the classic question about the Morning and the Evening Star), a methodology for normalizing, integrating, and clustering TA names presumably corresponding to the same entity. Using \texttt{HiP}, we analyze a large dataset collected from 15 sources and spanning 13,371 CTI reports, 17 vendor taxonomies, 3,287 TA names, and 8 mappings between them. Our analysis of the resulting name graph provides insights on key features of the problem, such as the concentration of aliases on a relatively small subset of TAs, the evolution of this phenomenon over the years, and the factors that could explain TA name proliferation. We also report errors in the mappings and methodological pitfalls that contribute to make certain TA name clusters larger than they should be, including the use of temporary names for activity clusters, the existence of common tools and infrastructure, and overlapping operations. We conclude with a discussion on the inherent difficulties to adopt a TA naming standard, a quest fundamentally hampered by the need to share highly-sensitive telemetry that is private to each CTI vendor.
\end{abstract}



\begin{keywords}
Cyber threat intelligence \sep Threat actor naming
\end{keywords}

\maketitle

\section{Introduction}\label{sec:introduction}

Hesperus is the name that the ancient Greeks gave to the Evening Star, a bright object that appears in the west after the Sun sets. 
Phosphorus is the name they gave to the Morning Star, which can be seen in the eastern morning sky shortly before the Sun rises.
At some point, ancient civilizations learned that these two celestial objects are one and the same ---the planet we now know as Venus.
In the philosophy of language, the sentence ``Hesperus is Phosphorus'' (HiP) refers to a famous puzzle often attributed to the philosopher and mathematician Gottlob Frege.
The HiP debate gained significant attention after the work of Saul Kripke in the 1970s \cite{Kripke80} and involves discussions about designation, reference of proper names, and the nature of identity \cite{Linsky59,Carney89}. 

The discipline of cyberthreat intelligence (CTI) has an endemic problem with nomenclature.
The issue goes back to the early days of the field, and one of the most representative examples is the multiplicity of names for malware families given by different AV vendors~\cite{Szor2005}. 
The Virus Naming Convention~\cite{CARONaming} proposed in the early 1990s by the Computer Anti-virus Research Organization's (CARO) constituted a solid attempt to standardize how antivirus companies name malware. The initiative was certainly influential and other similar efforts followed, such as the Common Malware Enumeration~\cite{Beck2006CME}, but to this day researchers need to resort to purposely created tools such as AVClass~\cite{SC2020} or Euphony~\cite{HSDBLKC17} to harmonize AV labels.

When it comes to attributing cyber attacks to state-sponsored threat actors (TAs) ---the so-called Advanced Persistent Threats (APTs)---, nomenclature is a HiP problem on steroids.
CTI vendors have developed their own analysis and attribution methodologies and, more importantly, their own naming schemes and catalogs for TAs~\cite{CrowdStrike20XXNaming, MITREATTACKGroups, Unit42Naming, MicrosoftThreatActorNaming}.
After two decades of APT reporting by dozens of CTI shops, the current CTI landscape is fragmented and plagued with a multiplicity of catalogs for naming TAs.
This problem creates the need for a resource that helps analysts to know whether the TA known as `X' by one vendor is the same as what another vendor names `Y'~\cite{Greenberg23,Poireault23}.
These mappings between TA names are key to enabling multiple applications, notably the creation of more accurate and complete TA profiles by combining intelligence labels and other content published by different sources.

Not having a system to unambiguously name TAs has practical negative consequences.
An established nomenclature is a sign of the maturity of a discipline.
Producing a worldwide system for naming objects requires a knowledge of their key features, structure and constituent elements, relationships among or within groups of objects, combination rules, and so on.
A good designator not only identifies a referent (the object it refers to), but also implicitly provides a mechanism for recognizing it and, in some cases, rules for automatically construct names for newly discovered objects.
When it comes to TAs and their activity, even if proper names are seen as mere tags with no meaning beyond their reference, designating the same object with different names leads to ambiguity.
In turn, this ambiguity could create uncertainty and confusion across incident response teams and security operation centers, resulting in delayed and sub-par responses.
The problem is arguably more severe if different observers use their own name to refer to the same object but they cannot trivially agree on whether they are talking about the same entity, because to do so they would have to share sensitive information with each other.
But sharing CTI is not always easy or possible at all.
In some cases, it may be beyond the interest of one of the parties.
In other cases, CTI shops must adhere to regulations concerning customer data protection or national security, which complicate intelligence sharing ---more so on a global scale across different regulatory environments.

This is a fundamentally difficult problem to solve because of the intrinsic nature of CTI products and processes. A typical workflow in a CTI shop involves collecting observations about an intrusion ---victims and their attributes, malware samples and tools used, Internet domains, and IP addresses controlled by the adversary--- and creating relationships among them. These relationships grow into activity clusters that eventually receive a designation for internal reference and, sometimes, also in published or privately shared reports. But these clusters do not necessarily refer to a single TA or a single operation. In some cases, multiple TAs reuse infrastructure, malware, and other resources even if their goals and targets are different. For example, Winnti is a modular remote access Trojan (RAT) that has been used by multiple TAs since 2010. The intrusion activity related to Winnti was initially grouped together and referred to as the Winnti Group~\cite{winnti_group}. Later, it was discovered to be multiple TAs using the same tool or related to the original group, such as Axiom, APT17, Ke3chang, and Aquatic Panda. Just by looking at an intrusion set, it may be impossible for the analyst to tell whether it corresponds to a single or multiple TAs and how they could relate to other known TAs, especially when TAs could split, merge, and change their TTPs over time. The bottom line is that the characterization (and designation) that a CTI vendor does of a specific TA activity could well be different from what other CTI vendor sees, even in the case that both have access to the evidence, which is not typically the case.

\subsection{Contributions}
CTI researchers have discussed the TA nomenclature problem for years~\cite{JAGS18, Roth18}.
To alleviate the problem, some key vendors have started very recently to publish one-to-one mappings between their catalogs~\cite{MicrosoftThreatActorNaming, Unit42TAgroups}.
However, the community currently lacks tools to map the TA taxonomies of major CTI vendors, a problem that has often been discussed as ``absurdly out of control''~\cite{Greenberg23,Poireault23}. More importantly, there is a general lack of understanding of the challenges and potential pitfalls that such an effort may entail. 

To fill this gap, in this work we present \toolname, a tool that generates clusters formed by the TA names given to the (presumably) same entity by different vendors. \toolname leverages public catalogs of TA names and mappings among catalogs to quantitatively study the proliferation of TA names and its causes. We use \toolname to analyze TA data from 15 sources, from which we extract 13,371 CTI reports, 3,287 TA names, 17 vendor taxonomies, and 8 mappings providing aliases between taxonomies. This large-scale corpus allows us to explore both quantitatively and qualitatively the challenges of mapping TA names at scale. The key findings of our analysis include:
\begin{itemize}
\item Processing multiple TA taxonomies and mappings requires a substantial revision and normalization effort. TA names contain issues ranging from misspellings due to human error, inconsistent use of naming rules (e.g., by using prefixes such as ``The'' or suffixes such as ``Group'' for the same TA), or an arbitrary use of case (lower/upper) and separators (spaces, dashes, and so on).

\item \toolname groups the 3,287 TA names into 977 alias clusters. The size distribution of this clustering is heavily skewed, with 62\% of the clusters having size 1 and 90\% of the clusters having size equal to or lower than 4. Only 2\% of the clusters have 15 aliases of more. These results indicate that the alias proliferation problem is concentrated on a relatively small number of entities according to the used mappings.

\item We quantitatively explore the extent to which alias proliferation is correlated with a number of TA features for which we have data in our sources. We find a strong positive linear correlation between the number of aliases of a TA and both the number of (public) reports on the actor and the number of CTI vendors reporting it. We also find a strong, though non-linear, relationship with the number of years a TA has been studied.

\item We study how the proliferation of TA names evolved over the years. Again, the results suggest that the phenomenon is concentrated on just a few TAs for which the number of aliases grew first in the period 2012-2018, and then in a second wave  after 2020.

\item We perform a manual inspection of the largest alias clusters with the aim of identifying systemic factors that contribute to the proliferation of names. Our findings include examples of human errors reporting TA equivalences that are not true. We also identify questionable aliases whose truth is impossible for us to verify. Finally, we discuss how the very nature of the CTI production process organically results in a variety of designators connected to each other, including temporary names for TAs, tools, infrastructure, overlapping operations, and known adversary groups. This is not a bug, but a feature of how attribution works in practice ---and one that significantly contributes to the number of aliases of certain TAs.

\item We conclude with a discussion on the possibility of reaching consensus on a universal TA naming standard. In our opinion, this quest might be possible in theory but very hard to adopt without first addressing the fundamental problem of sharing telemetry data, which is private to each CTI vendor and potentially very sensitive. In the meantime, better curated mappings and tools such as \toolname could be marginally useful for certain applications.
\end{itemize}

\paragraph{Research artifacts}
We open source \toolname and the datasets (raw source data, vendor taxonomies, attributed mappings, and name clusters) used in the analysis presented in this work.\footnote{\url{https://github.com/0xjet/hip}}

\section{HiP: Mapping threat actor names from multiple taxonomies}
\label{sec:hip}
In this section, we describe a tool named \toolname (Hesperus is Phosphorus) that normalizes and integrates the existing public taxonomies of TA names and the mappings between them. \toolname allows us to perform a quantitative analysis of the TA naming problem in order to understand its extent, scope, and the factors that contribute to the proliferation of aliases.

\begin{figure*}[h!t]
\centering
\includegraphics[width=\linewidth]{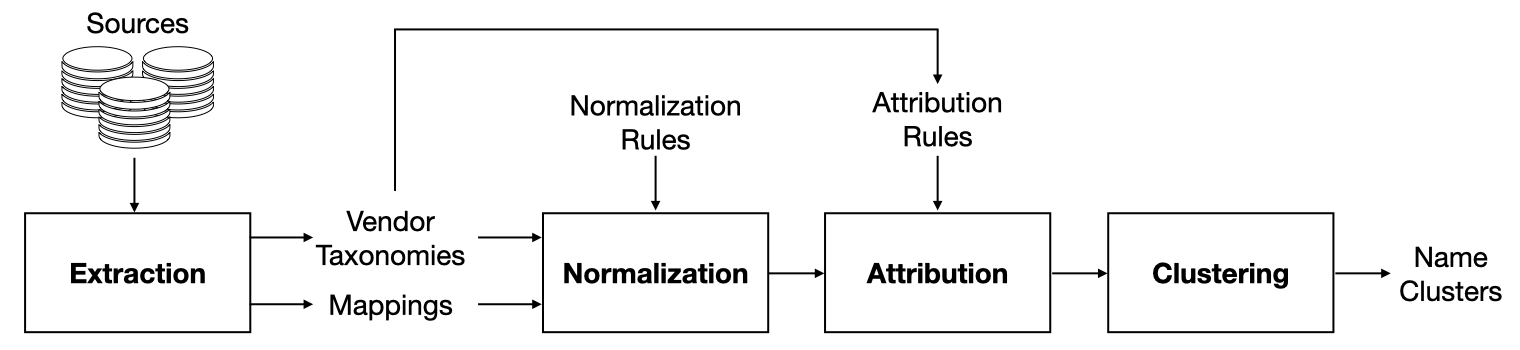}
\caption{\toolname architecture.}
\label{fig:hiparch}
\end{figure*}

The architecture of \toolname is shown in Figure~\ref{fig:hiparch}. The tool feeds from a number of official and community-driven sources that provide listings of TA names. These names are typically linked to other CTI data such as TA profiles, activity reports, and malware families. They often also contain a short list of aliases (i.e., mappings) for each TA name. We generate a set of vendor taxonomies and a set of name mappings from these sources, which are further processed to produce a collection of name clusters. Each name cluster is a graph in which nodes are normalized and attributed TA names, and edges represent an alias relationship given by a mapping. The name clusters can be queried through a search API that returns the cluster of aliases of a given TA name.

The remainder of this section describes the processing pipeline of \toolname in detail.

\subsection{Extraction}

\begin{table*}[h!t]
\centering
\caption{Current data sources used in \toolname.}
\label{tab:sources}
\begin{tabular}{lp{13cm}}
\toprule
\textbf{Source} & \textbf{Description}\\
\midrule
APTGO~\cite{threatActorSpreadsheet} &
The APT Groups and Operations is a community-driven knowledge base compiling multiple vendor taxonomies and mappings between them.\\
Malpedia~\cite{malpedia} &
Repository of malware families and related CTI maintained by Fraunhofer FKIE. It offers standardized metadata, including TA aliases, associations to malware families, and links to reports.\\
Microsoft Mapping~\cite{MicrosoftThreatActorNaming} &
Microsoft's official description of its TA taxonomy and some known aliases for each threat actor.\\
MISP Galaxy Clusters~\cite{mispThreatActorGalaxy} &
Multiple clusters contained in the MISP Galaxy focused on TA names and their aliases.\\
ORKL~\cite{orkl2025} &
Large collection of open-source CTI reports annotated with relevant metadata, including TA names and aliases. ORKL aggregates data from 11 distinct sources, including MITRE ATT\&CK data~\cite{mitre_attack_stix}, APTnotes~\cite{aptnotes_data}, CyberMonitor's APT \& Cybercriminals Campaign Collection~\cite{cybermonitor_apt},  ETDA's Threat Group Cards~\cite{etda_apt_groups}, Alienvault's Open Threat Exchange (OTX)~\cite{alienvault_otx}, Secureworks Threat Profiles~\cite{secureworks_threat_profiles}, ORKL Community Contributed Content~\cite{orkl2025}, and VX Underground~\cite{vx_underground}\\
\bottomrule
\end{tabular}
\end{table*}

\paragraph{Sources}
Unfortunately, most CTI vendors do not publish structured catalogs of their TA naming taxonomies. Instead, the names are often derived from collections of published reports or shared CTI items. \toolname~relies on 12(+3) well-known structured data sources to derive a comprehensive collection of vendor taxonomies and mappings (see Table~\ref{tab:sources}). The largest source is ORKL~\cite{orkl2025}, a large collection of structured CTI metadata linked to 13,371 reports that are integrated from 11 relevant sources. We independently crawl data from MISP~\cite{mispThreatActorGalaxy}, Malpedia~\cite{malpedia}, and the APT Groups and Operations (APTGO) repository~\cite{threatActorSpreadsheet} because we noticed the content from these three sources in ORKL was not up to date.

\paragraph{Taxonomies}
Each source is automatically processed to extract embedded taxonomies and mappings. We create one taxonomy per known vendor and codify rules to recognize names that belong to it. For example, if a TA name has the format \texttt{G[DDDD]}, where \texttt{D} are digits, it is added to the MITRE taxonomy. We keep a special taxonomy named \texttt{UNK} (unknown) for TA names that cannot be assigned to a known vendor taxonomy. The process is based on recognizable naming rules extracted from a list of known CTI vendors. This process is not fully automated, since many taxonomies follow name patterns that are hard to codify. To account for this limitation, we manually inspect the resulting taxonomies to correct errors. In total, we extract 3,287 TA names, 1307 of which are associated with 17 known vendor taxonomies. Table~\ref{tab:taxonomies} lists the set of taxonomies extracted from the sources and their size.

\begin{table}[h!t]
\centering
\caption{Extracted taxonomies and their sizes (number of TA names).}
\label{tab:taxonomies}
\begin{tabular}{lr}
\toprule
\textbf{Vendor} & \textbf{Taxonomy size}\\
\midrule
UNK  & 1,980\\
CrowdStrike  & 241\\
Microsoft  & 200\\
Secureworks  & 193\\
Mandiant  & 155\\
Microsoft (old)  & 123\\
MITRE  & 110\\
Palo Alto Unit 42  & 71\\
Thales  & 47\\
CERT-UA  & 35\\
Recorded Future & 32\\
360  & 27\\
FireEye  & 19\\
Kaspersky  & 17\\
Symantec  & 14\\
Cisco Talos  & 14\\
Tencent  & 7\\
NSA  & 2\\
\midrule
Total & 3,287\\
\bottomrule
\end{tabular}
\end{table}

\paragraph{Mappings}
Some sources provide mappings that associate a canonical TA name with a number of aliases (i.e., other names for the TA). These mappings typically do not indicate the vendor or taxonomy of each name in the list of aliases. Table~\ref{tab:mapping} lists all the mappings used in \toolname~along with their size and the average and maximum number of aliases per TA name.

\begin{table}[h!t]
\centering
\caption{Extracted mappings and their sizes.}
\label{tab:mapping}
\begin{tabular}{llrrrrr}
\toprule
\textbf{Mapping} & \textbf{Source} & \textbf{Size} & \textbf{Avg.} & \textbf{Max.} \\
\midrule
APTGO & APTGO~\cite{threatActorSpreadsheet}  & 237  & 4.24 & 26\\
Paloalto & Paloalto~\cite{Unit42TAgroups}  & 50  & 4.34 & 17\\
Malpedia & Malpedia~\cite{malpedia}  & 314  & 3.85 & 41\\
ETDA & ORKL~\cite{orkl2025}  & 270  & 4.98 & 46\\
MITRE  & ORKL~\cite{orkl2025}  & 175  & 2.93 & 15\\
Secureworks  & ORKL~\cite{orkl2025}  & 106  & 4.75 & 17\\
MISP Galaxy & MISP Galaxy~\cite{mispThreatActorGalaxy}  & 341  & 3.55 & 39\\
Microsoft & Microsoft~\cite{MicrosoftThreatActorNaming}  & 114  & 3.64 & 14\\
\bottomrule
\end{tabular}
\end{table}

\subsection{Normalization}
While inspecting the taxonomies and mappings extracted from the sources, we noticed multiple naming inconsistencies and errors that, if not properly handled, would result in an inaccurate integration of the data. The main issues we identified are the following.
\begin{enumerate}

\item\textit{Misspellings}, likely attributable to human typing errors during transcription. Examples include \texttt{Agoniznig Serpens} vs. \texttt{Agonizing Serpens}, \texttt{Calisto} vs. \texttt{Callisto}, \texttt{Kimsuki} vs. \texttt{Kimsuky}, \texttt{Red Bald Knight} vs. \texttt{Red Bald Night}, \texttt{Qudedagh} vs. \texttt{Quedagh}, \texttt{Nemim} vs. \texttt{Nemin}, and many others.

\item\textit{Group variants} referring to the same TA name. A recurrent pattern consists of appending generic collective terms such as ``Group'', ``Lair'', ``Gang'', ``APT'' or ``Cyberespionage'' to the canonical TA name. Examples include \texttt{Black Energy} vs. \texttt{Black Energy Group} vs. \texttt{Black Energy (Group)}, \texttt{Budminer} vs. \texttt{Budminer Cyberespionage Group}, and \texttt{Callisto} vs. \texttt{Callisto Group}.

\item\textit{Use of prefixes}, notably ``The'' before the canonical name. Examples include \texttt{The Gorgon Group} vs. \texttt{Gorgon Group}, \texttt{The Lamberts} vs. \texttt{Lamberts}, \texttt{The Mask} vs. \texttt{Mask}, and \texttt{The Shadow Brokers} vs. \texttt{Shadow Brokers}.

\item\textit{Vendor suffix}. We identified a class of designations in which the TA name is followed, generally in parentheses, by the associated CTI vendor. Examples include \texttt{APTC38} vs. \texttt{APTC38 (Qianxin)}, \texttt{Earth Preta} vs. \texttt{Earth Preta (Trendmicro)}, \texttt{Sharpdragon} vs. \texttt{Sharpdragon (CHKPT)}, and \texttt{Hive 0081} vs. \texttt{Hive 0081 (IBM)}.

\item\textit{Inconsistent use of case and separators}, such as spaces, dashes, underscores, slashes, and line breaks. Examples include \texttt{scarred\_manticore} vs. \texttt{Scarred Manticore}, \texttt{Sands Casino} vs. \texttt{sands\_casino}, \texttt{apt32} vs. \texttt{APT 32} vs. \texttt{APT-32}, and \texttt{scarleteel} vs. \texttt{SCARLETEEL} vs. \texttt{ScarletEel}.

\end{enumerate}

To address these issues, we manually developed a set of extensible normalization rules that can be augmented as additional inconsistencies are observed. To facilitate matching of the resulting names, we adopt the convention of removing all spaces, dashes, underscore symbols, slash symbols, and line breaks. Finally, the resulting string is converted to upper case.

\subsection{Vendor attribution}
We attribute each normalized TA name to a vendor to produce attributed names of the form \texttt{VENDOR:NAME}. This is a two-step process that begins with a  rewriting of the taxonomies and mappings extracted from the sources using the normalized names derived in the previous step. We then search for the presence of each name in the vendor taxonomy, renaming it as \texttt{VENDOR:NAME}. Again, we use a special vendor designation (\texttt{UNK}) for names that cannot be attributed. The output of this process is a set of attributed mappings with normalized names.

\subsection{Clustering}

The final step in the pipeline of \toolname takes as input the attributed mappings and constructs a graph $G=(V,E)$ representing TA names and their aliases, where:
\begin{itemize}
\item $V$ is a set of vertices, where each vertex $v \in V$ represents the normalized and attributed name of a TA, in the form \texttt{VENDOR:NAME}.
\item $E \subseteq V \times V$ is a set of edges, where an edge $e=(v_i, v_j) \in E$ exists iff $v_i$ is included in the list of aliases of $v_j$ in an attributed mapping. We also characterize each edge by a confidence value that measures how many mappings report $v_i$ as an alias of $v_j$. To track alias relationships, we also store the mapping for each edge.
\end{itemize}

We refer to this graph as the Threat Actor Name Alias Graph (TANAG). The TANAG is made up of a collection of disconnected subgraphs, where each connected component represents a (presumably) unique entity with multiple names according to multiple vendors and mappings. We refer to each connected component of the TANAG as a \emph{name} or \emph{alias cluster}.

\section{The threat actor naming mess quantified}
\label{sec:quant}

This section presents the result of applying the methodology described in Section~\ref{sec:hip}. It then studies the impact of different factors that could contribute to the proliferation of TA names and examines its evolution over time.

\begin{figure*}[h!t]
\centering
\includegraphics[width=\linewidth]{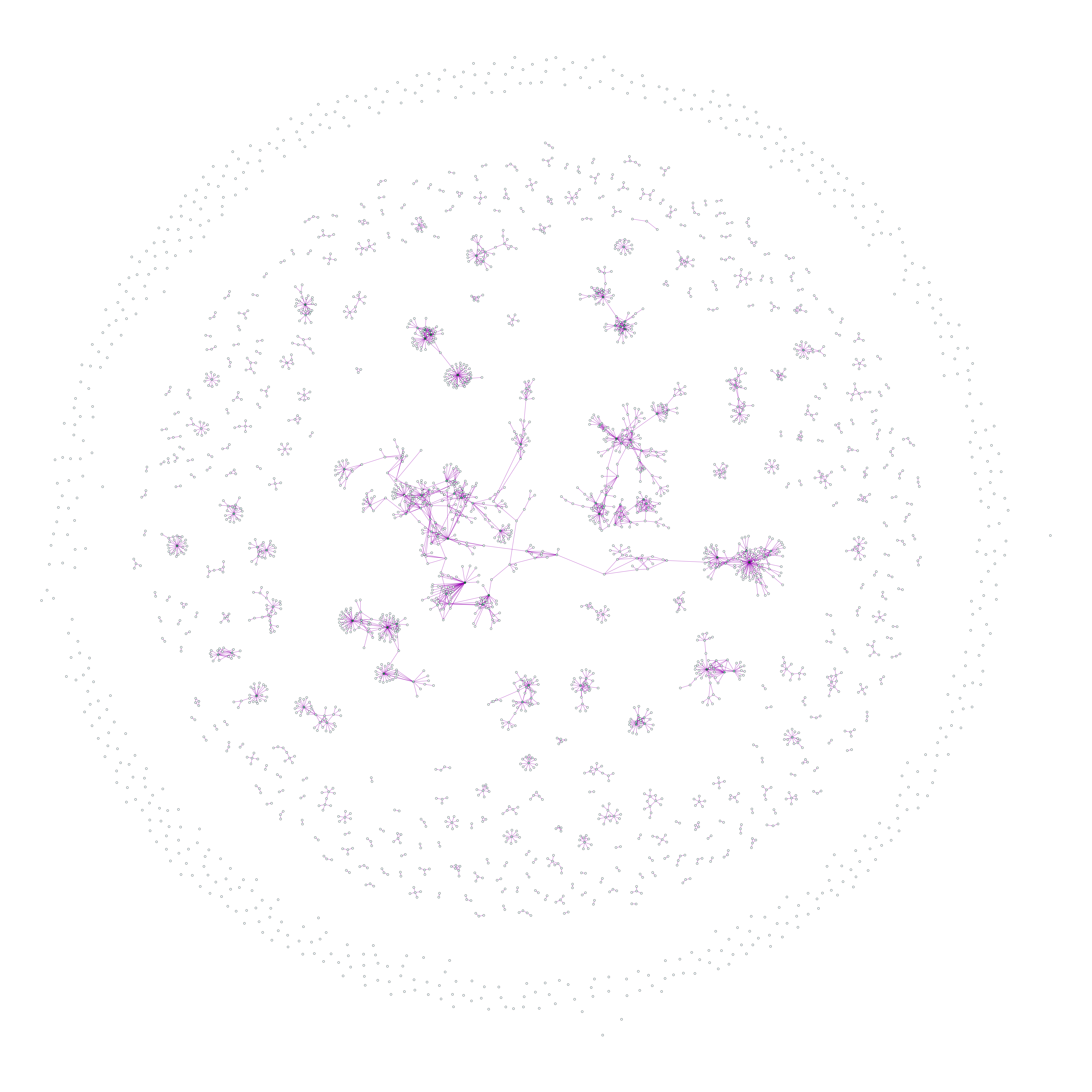}
\caption{Threat Actor Name Alias Graph (TANAG) produced by \toolname.}
\label{fig:ta_naming_graph}
\end{figure*}

\subsection{TANAG}

The resulting TANAG contains 3,287 vertices grouped into 977 connected components (name clusters). Figure~\ref{fig:ta_naming_graph} shows the complete graph, with most of the alias clusters of size one located in the external ring.\footnote{An interactive explorer for the TANAG is availabe at: \url{https://0xjet.github.io/tanag.html}.}
Next, we discuss the key properties of this graph.

\begin{figure}
\centering
\includegraphics[width=\columnwidth]{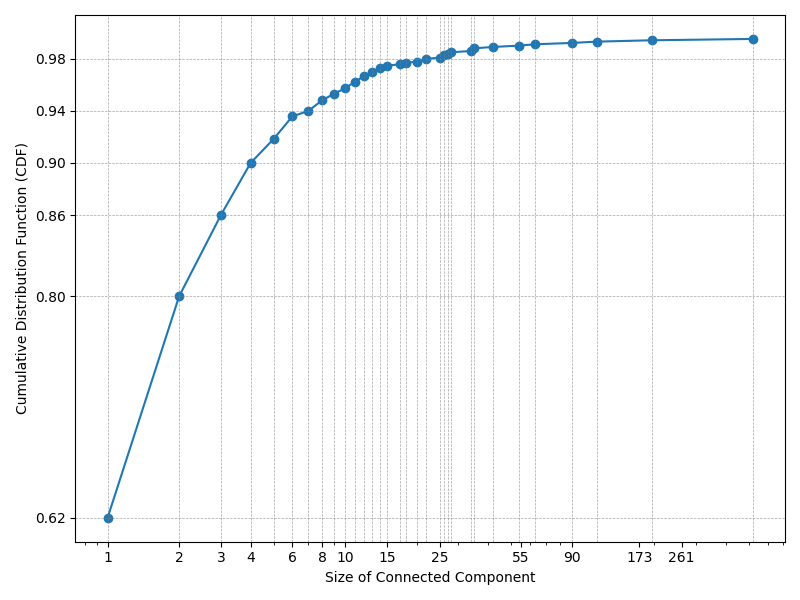}
\caption{Cumulative distribution function of the TA alias cluster sizes.}
\label{fig:ac_cdf}
\end{figure}

\paragraph{Distribution of the alias cluster size.}
Figure~\ref{fig:ac_cdf} shows the cumulative distribution function of the alias cluster sizes. The average and median sizes of a cluster are 1 and 3.42, respectively, but the distribution is heavily skewed. Up to 62\% of the alias clusters have size 1, and 90\% of the clusters have size equal to or lower than 4. Only 2\% of the clusters have 15 aliases or more. These results indicate that most (>90\%) TAs receive either a single name or a very reduced (<4) number of aliases. Thus, the name proliferation phenomenon either concentrates on a very small number of TAs, or the fragmentation of visibility across vendors results in multiple names to the same actor. The larger alias cluster contains 520 aliases\footnote{Including APT 9 APT 15, APT 17, APT 19, APT 26, APT 41, AXIOM, BACKDOOR DIPLOMACY, BARIUM, BEIJING GROUP, BRONZE ATLAS, BRONZE EXPORT, BRONZE FIRESTONE, BRONZE KEYSTONE, BRONZE PALACE, DEEP PANDA, ELDERWOOD, HIDDEN LYNX, KE3CHANG, NIGHTSHADE PANDA, TURBINE PANDA, WINNTI GROUP.} and corresponds to a cluster of TAs historically associated with China's PLA.

\paragraph{Intelligence gain: The case of malware families}
To better understand the relevance derived from combining existing TA names into alias clusters, we quantify the gain resulting from merging relevant attributes of the TA profile. For this analysis, we use the number $N_m(t_i)$ of malware families associated to each TA $t_i$, as listed in the Malpedia source. We first aggregate at the alias cluster level all malware families linked to the TAs belonging the cluster. Let $N_m(C_j)$ be the number of malware families associated with the alias cluster $C_j$. We then compute for each TA $t_i \in C_j$ a Malware Intelligence Gain (MIG) score as the ratio between the number of malware families in its profile and the number of malware families in its alias cluster, i.e.:
$$\mathrm{MIG}(t_i) = \frac{N_m(C_j)}{N_m(t_i)}$$
The MIG captures how much broader the intelligence about malware families used is in the alias cluster compared to each of its individual members. A value of 1 indicates that $t_i$ already covers all families in the alias cluster and, therefore, no new intelligence is gained by merging its profile with those of known aliases. while values greater than 1 indicate that the alias cluster provides additional intelligence beyond what the individual profile gives.

Unfortunately, we cannot compute the MIG at the individual TA level. Our source, Malpedia, provides only a limited mapping between aliases, listing malware families for a set of known aliases belonging to the same TA. Since these sets of aliases are subclusters of each larger cluster in our TANAG, we compute the MIG for each of these subclusters by extending the previous definition to account for the malware families of each set of aliases. We then analyze the MIG value with respect to the subcluster's size relative to the alias cluster to which it belongs.

\begin{figure}
\centering
\includegraphics[width=\columnwidth]{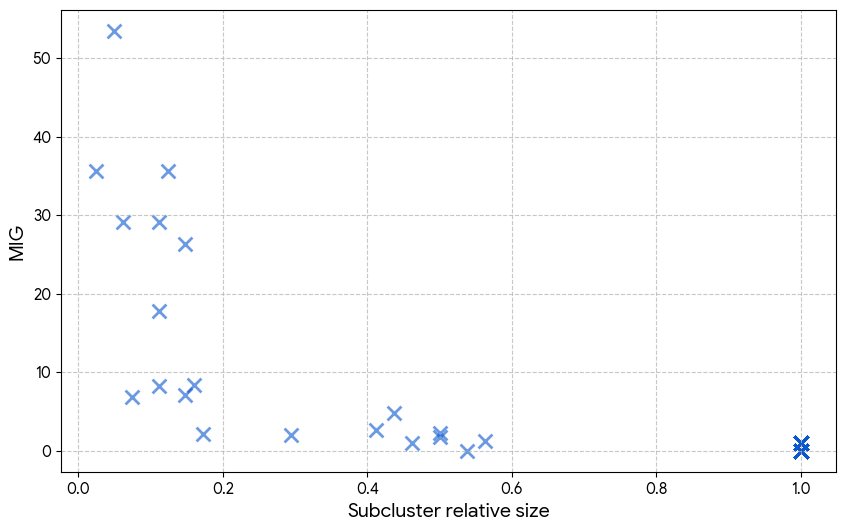}
\caption{Malware Intelligence Gain (MIG) vs. subcluster size.}
\label{fig:cdf_mig}
\end{figure}

The results are presented in Figure~\ref{fig:cdf_mig}. As expected, the gain for smaller subclusters is significantly large, with numerous examples of subclusters having an MIG greater than 20-30 for subclusters with a relative size inferior to 20\% of the alias cluster. The highest MIG case we identify is a subcluster containing the TAs \texttt{BASIN}, \texttt{TA416}, \texttt{G0129}, and \texttt{PKPLUG}, which are collectively linked to six malware families. These four TA names are integrated into a larger cluster containing 81 aliases and 321 malware families, resulting in a MIG of 53.5. As the relative size of the subcluster increases, the MIG value tends to remain low, typically ranging between 1 and 8 for subclusters between 30\% and 60\% of the parent cluster's size. Finally, there is no gain (MIG=1) for large clusters that approach the size of the parent cluster.

\subsection{Explaining TA name proliferation: Correlation analysis}
\label{sec:corr_analysis}

We explore the extent to which the proliferation of aliases could be explained by the following variables (features):
\begin{itemize}
\item $n_r =$ Number of CTI reports in which any of the aliases in the cluster appear.
\item $n_v =$ Number of CTI vendors that have published a CTI report about the TA.
\item $n_y =$ Number of years that the TA has been studied (by any CTI vendor).
\item $n_g =$ Number of different geographies or world regions targeted by the TA.
\item $n_s =$ Number of victim sectors targeted by the TA.
\end{itemize}

Some of these features are directly available in the TA profile metadata, while others require processing of the datasets of CTI reports. In total, we extract 15,944 records from 5 sources contained in ORKL (13,340 records) and MISP (2,604 records). These records provide labeled data spanning 26 years (2000 through 2025), 9 CTI vendors, 66 sectors and 170 geographies.

\begin{figure}
\centering
\includegraphics[width=\columnwidth]{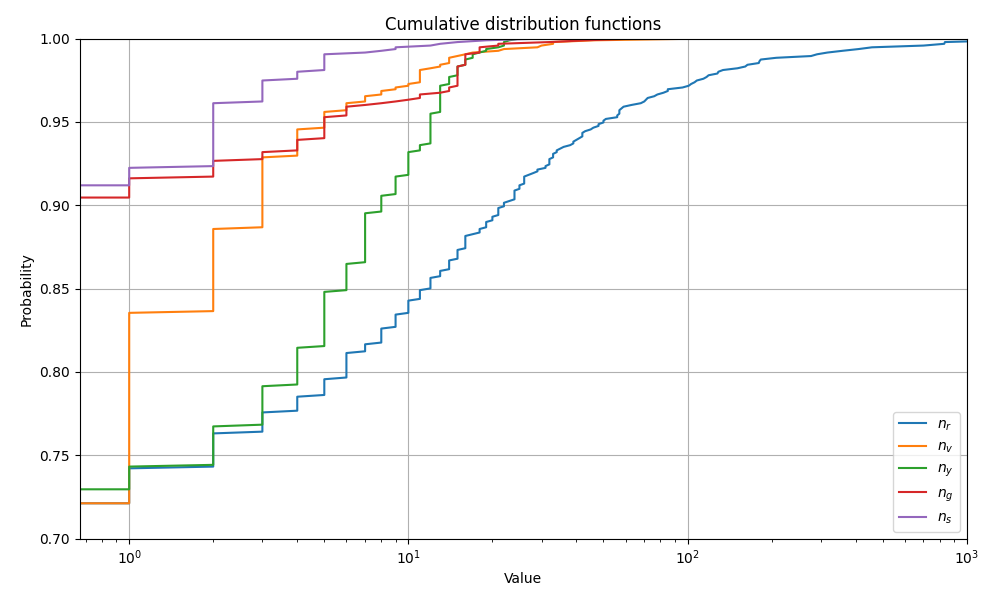}
\caption{Cumulative distribution function of the studied TA features vs. the size of the TA alias cluster.}
\label{fig:ac_features_cdf}
\end{figure}

We first compute a feature vector for each unique TA name and then aggregate them to derive a feature vector for each alias cluster. Figure~\ref{fig:ac_features_cdf} shows the distribution of each variable with respect to the size of the alias cluster. The number of reports ($n_r$) seems to have a larger spread, and the crossings between some of them indicate that their distributions differ in shape, not just in central tendency. The number of attacked sectors ($n_s$) and geographies ($n_g$) have the lowest variance.

We compute Pearson's correlation coefficient to measure linear correlation. We also compute Spearman's rank correlation coefficient and Kendall's Tau to capture monotonic correlation that could be nonlinear. Table~\ref{tab:corr_analysis} shows the correlation coefficients and p-values. The linear correlation plots and cross-correlation values between features are shown in Figure~\ref{fig:ac_cdf}.

\begin{table*}
\caption{Correlation analysis of studied TA features vs. the size of alias clusters}
\label{tab:corr_analysis}
\resizebox{\linewidth}{!}{%
\begin{tabular}{lllll}
\toprule
TA feature & Pearson & Spearman & Kendall & Interpretation\\
\midrule
$n_r$ & 0.92 (p=0) & 0.83 (p=6.59e-249) & 0.77 (p=5.72e-163) & Very strong, highly linear\\
$n_v$ & 0.88 (p=1.94e-311) & 0.83 (p=1.13e-246) & 0.77 (p=1.11e-159) & Very strong, highly linear\\
$n_y$ & 0.35 (p=2.47e-29) & 0.82 (p=3.40e-231) & 0.75 (p=1.24e-151) & Strong, non-linear\\
$n_g$ & 0.70 (p=3.41e-138) & 0.29 (p=1.66e-19) & 0.27 (p=3.20e-19) & Inconclusive\\
$n_s$ & 0.71 (p=3.86e-149) & 0.27 (p=2.27e-17) & 0.25 (p=7.06e-17) & Inconclusive\\
\bottomrule
\end{tabular}
}
\end{table*}

\begin{figure*}[h!t]
\centering
\includegraphics[width=\linewidth]{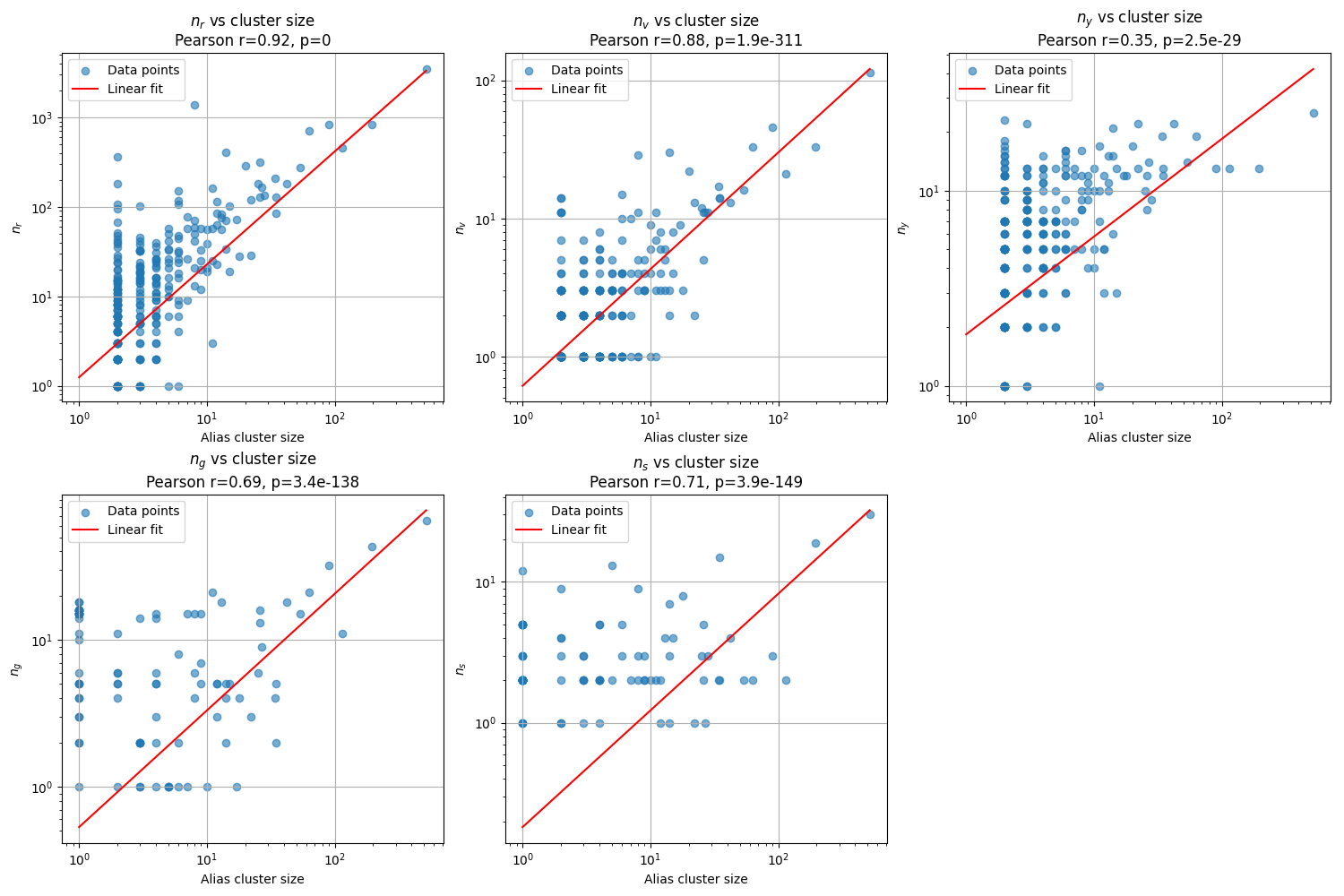}
\caption{Linear (Pearson) correlation plots and coefficient between TA features and alias cluster sizes.}
\label{fig:ac_corr_plots}
\end{figure*}

All correlations are statistically significant because of the extremely small p-values. The key conclusions from the correlation analysis are:
\begin{itemize}
\item The number of reports ($n_r$) and vendors ($n_v$) exhibit a very strong, statistically significant, and highly linear positive relationship with the size of the name clusters. The close agreement between the Pearson and the rank-based (Spearman, Kendall) coefficients suggests that the relationship is strongly linear and that the data do not contain influential outliers.
\item The number of years ($n_y$) exhibits a strong non-linear relationship. The low Pearson coefficient (0.35) indicates a poor linear fit, but the very high Spearman and Kendall coefficients (>0.75) suggest a consistent monotonic relationship.
\item The number of geographies ($n_g$) and sectors ($n_s$) attacked show a statistically significant and relatively strong linear relationship, with a Pearson coefficient of around 0.7 in both cases. However, the fact that the rank correlations are very weak ($\approx 0.26$ and $0.28$) indicates that the underlying directional tendency for the majority of the data is weak. This discrepancy strongly suggests the presence of extreme outliers that are driving the high Pearson coefficient, making the linear model appear much better than it truly is.
\end{itemize}

\subsection{Longitudinal analysis}
\begin{figure*}[h!t]
\centering
\includegraphics[width=\linewidth]{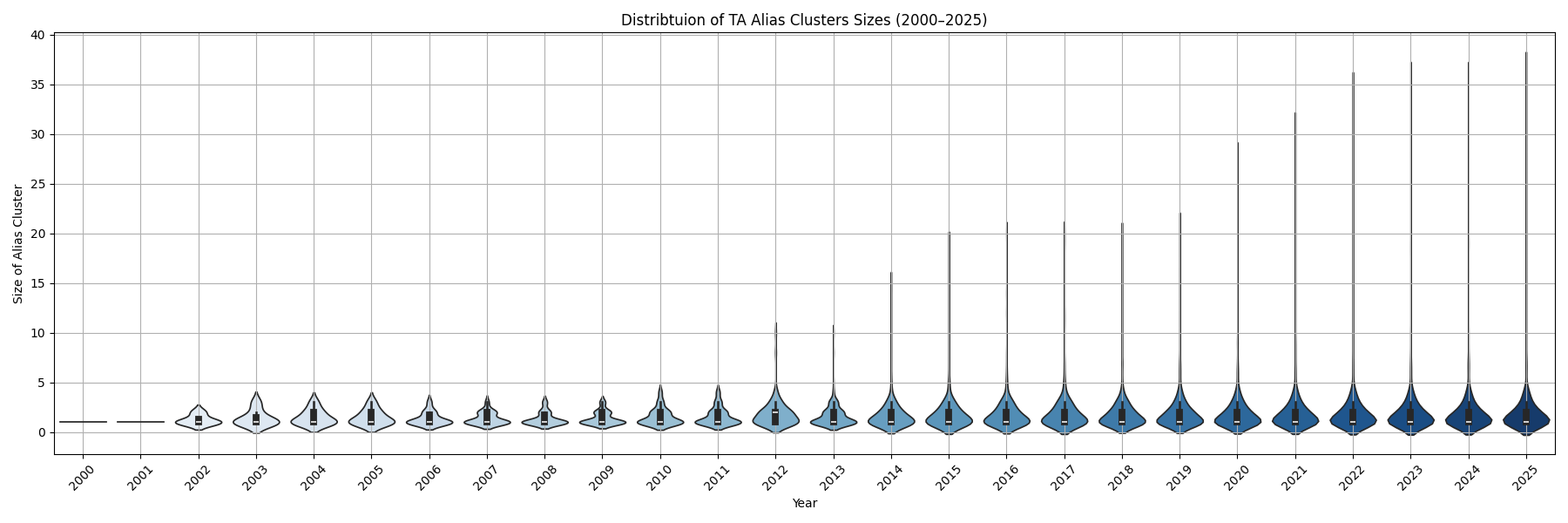}
\caption{Evolution (2000-2025) of the distribution of the TA alias cluster sizes.}
\label{fig:longitud}
\end{figure*}

We study how the proliferation of TA names evolved over the years. For this task, we rely on the same subset of data used in Section~\ref{sec:corr_analysis}, since these sources are the only ones providing timestamped labels (years) for each record. The oldest reports and TA profiles in the dataset are from the year 2000. We create 26 sub-datasets, $D_i, i=0, \ldots, 25$, where each $D_i$ contains the reports and TA profiles from year 2000 to 2000$+i$. We then obtain the TANAG for each dataset $D_i$ and the corresponding distribution of alias clusters. Figure~\ref{fig:longitud} shows all distributions longitudinally.

The results reveal that the number of aliases have increased over time, but the phenomenon is significantly focused on just a few TAs. The number of aliases for the majority of TAs has only increased from 1 in the early 2000s to less than 3 after 25 years. Between 2012 and 2018, a very reduced number of TAs saw their number of aliases grow to the range 10-20, and then to more than 30 in subsequent years. This could be a consequence of certain groups being more active, or their activity being more visible to CTI vendors. It could also be a limitation of current catalog mappings, which do not establish the equivalence of names that in reality correspond to the same TA.

\section{Errors, pitfalls, and other factors contributing to the proliferation of aliases}
\label{sec:errors}

The graph resulting from the aggregation of TA mappings across multiple vendors using \toolname is not necessarily an accurate representation of the ground truth of TA identities. Rather, it is the direct consequence of taking all public mappings at face value and combining them, reflecting the way (open-source) CTI has evolved over time. In addition to the quantitative analysis discussed in the previous section, we performed a manual inspection of the TANAG with the aim of identifying industry practices that contributed to the proliferation of names, and also errors in the data. Our key findings include:
\begin{enumerate}
\item Temporary names, overlapping operations, and long-standing ambiguities between actor definitions lead to redundant or conflicting labels. We find that even well-established clusters include a multiplicity of names within the same vendor over time (intra-vendor inflation). Over the years, a single vendor may naturally increase the number of aliases associated with a TA due to changes in internal tracking, the emergence of new information, or the adoption of updated classification schemes. This phenomenon reflects the organic and decentralized nature of threat intelligence production and naturally leads to name inflation.
\item In some extreme cases, name clusters include both tools and groups under the same alias, such as mixing infrastructure or malware families with operator names (e.g., the broad \texttt{Winnti} label). This is not necessarily due to analytical error, but rather an artifact of the industry's naming dynamics, where overlapping campaigns, unclear attribution boundaries, and evolving intelligence contribute to alias proliferation. The fact that a single vendor such as Mandiant uses multiple names linked to each other, such as \texttt{MANDIANT:UNC2630}, \texttt{MANDIANT:TEMP.BOTTLE}, and \texttt{MANDIANT:APT5} is not a bug, but a feature of how attribution works in practice.
\item Inter-vendor synonyms, as captured by taxonomy mappings, can contain incorrect relationships due to imperfect information or human errors. We find some examples of such errors, which in some cases are rooted in an unfortunate choice of names for an actor, tool, or operation that collides with an existing one.
\end{enumerate}

We next illustrate through various case studies some of the practices that lead to erroneous or questionable alias associations.

\subsection{Sloppy mappings: The case of Grizzly Steppe and APT29}

\begin{figure}
\centering
\includegraphics[width=\columnwidth]{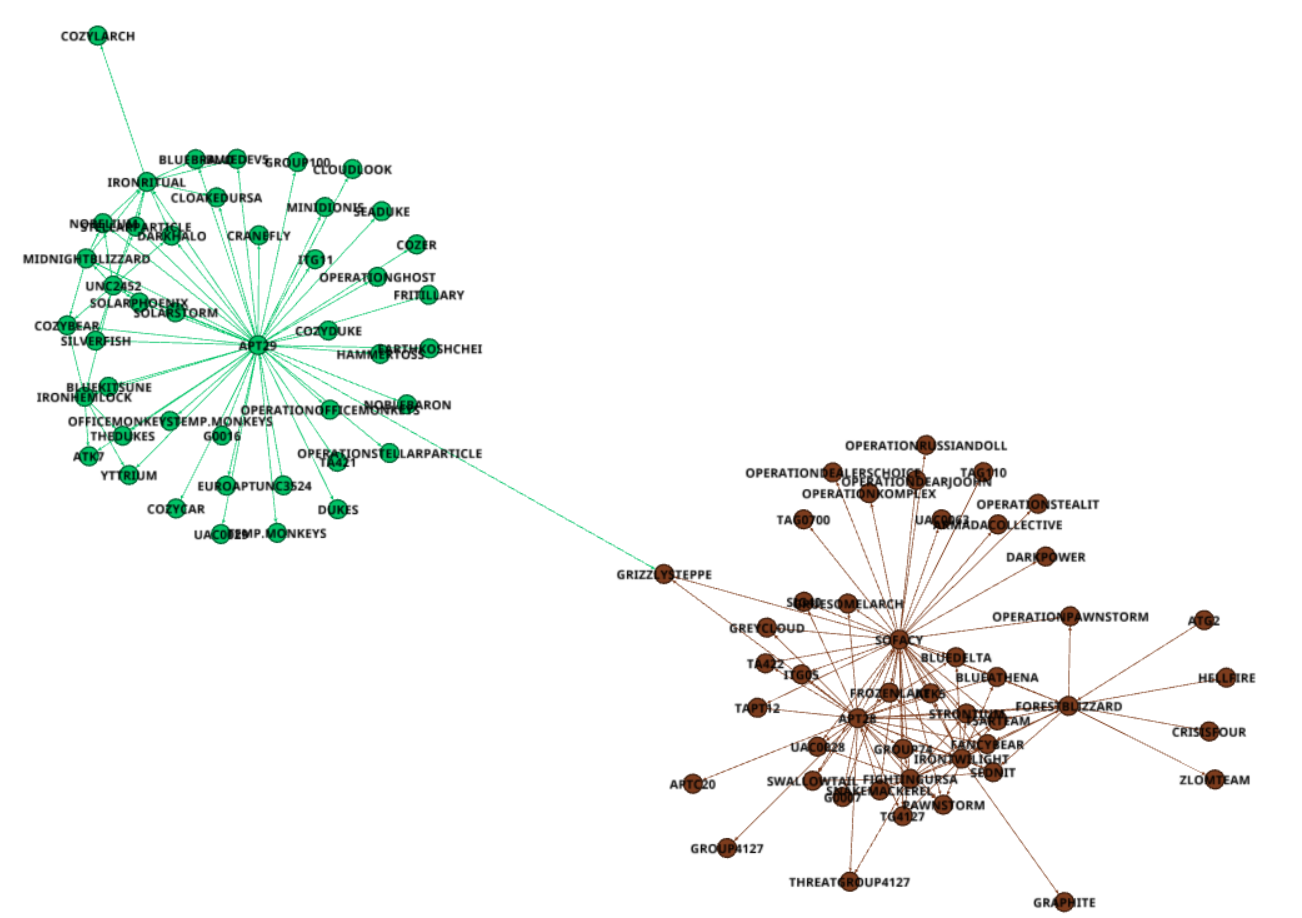}
\caption{The Grizzly Steppe vs. APT29 vs. APT28 cluster}
\label{fig:grizzly_steppe_and_apt29}
\end{figure}

Figure \ref{fig:grizzly_steppe_and_apt29} shows an alias cluster with a salient feature: There is a single edge connecting Grizzly Steppe (\texttt{UNK:GRIZZLYSTEPPE}) and APT29 (\texttt{MANDIANT:APT29}). That edge is the only connection bridging two clusters otherwise greatly interconnected. The weight of the edge is 1, meaning that there is only one source (MISP) claiming that both names are aliases. Additionally, we can see that Grizzly Steppe is connected to \texttt{MANDIANT:APT29} and \texttt{MANDIANT:APT28} nodes. Both names are given by the same CTI Vendor: Mandiant. However, it is not Mandiant who determines the association between both APT groups despite the fact that Mandiant is, allegedly, who possesses more detailed intelligence about APT28 and APT29. This connection seems to be an error whose origin can be traced to a CISA report~\cite{cisa} describing Grizzly Steppe as the name given to an operation associated to both APT groups:
\begin{quote}
\textit{``The U.S. Government confirms that two different RIS actors participated in the intrusion into a U.S. political party. The first actor group, known as Advanced Persistent Threat (APT) 29, entered into the party’s systems in summer 2015, while the second, known as APT28, entered in spring 2016''}.
\end{quote}

\subsection{Name collisions: The Two Operations Daybreak}
\begin{figure}
\centering
\includegraphics[width=\columnwidth]{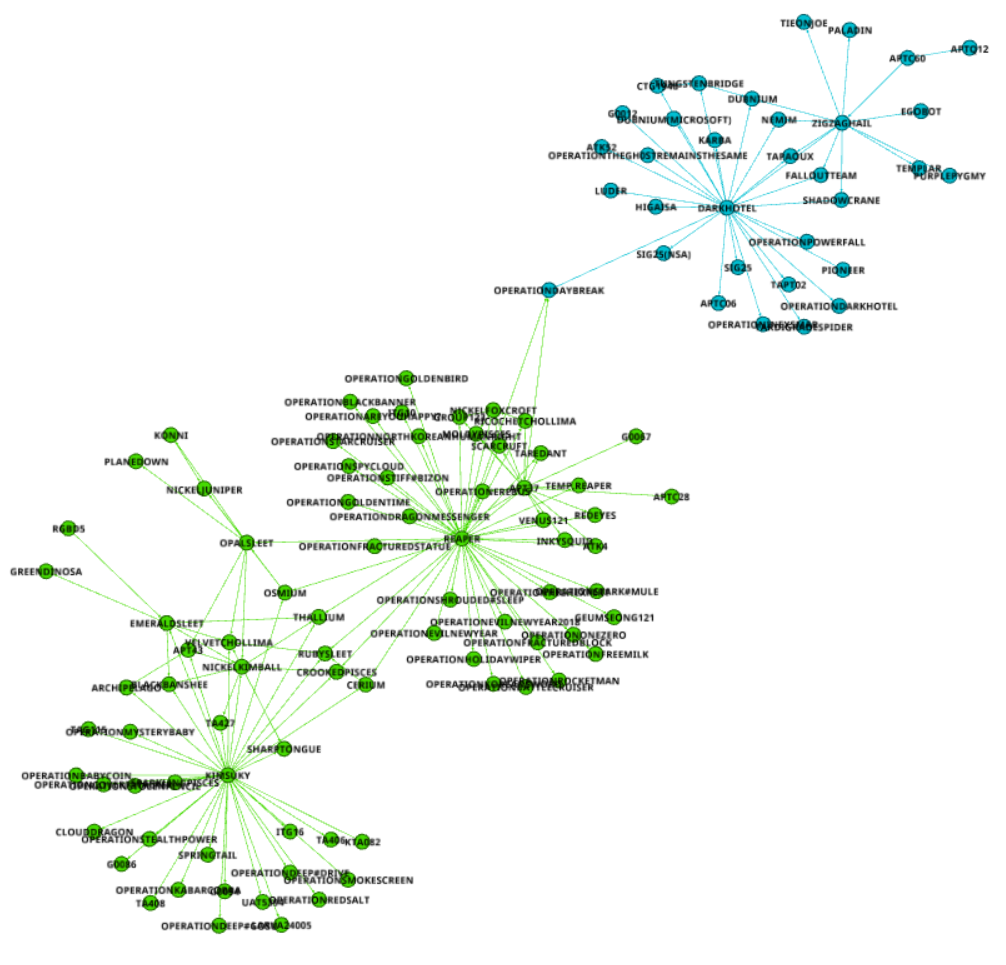}
\caption{The two Operation Daybreak cluster.}
\label{fig:operation_daybreak}
\end{figure}

The cluster shown in Figure~\ref{fig:operation_daybreak} contains two clearly delimited subclusters connected by a node labeled \textit{Operation Daybreak}. To one of the sides, this node connects to \texttt{MANDIANT:APT37} and \texttt{UNK:REAPER}, both of which are linked to many other aliases. To the other side, \texttt{UNK:OPERATIONDAYBREAK} connects to \texttt{UNK:DARKHOTEL}, which holds a central position in that subcluster. The weight of the link between Operation Daybreak and DarkHotel is 1, indicating that a single source (ORKL) establishes an alias relationship. This linkage traces to ETDA’s DarkHotel profile, whose list of operations performed includes the following entry dated December 2015~\cite{etdaDarkHotelThreat}:
\begin{quote}
\textit{``Operation Daybreak Method: Uses Flash zero-day exploit for CVE-2015-8651. Note: not the same operation as Reaper, APT 37, Ricochet Chollima, ScarCruft’s Operation Daybreak}.
\end{quote}
Likewise, in ETDA's Reaper/APT37 profile we can see the following entry in \emph{Operations Performed} dated in March 2016:
\begin{quote}
\textit{``Operation Daybreak. Target: High profile victims. Method: Previously unknown (0-day) Adobe Flash Player exploit. It is also possible that the group deployed another zero day exploit, CVE-2016-0147, which was patched in April. <https://securelist.com/operation-daybreak/75100/> Note: not the same operation as DarkHotel’s Operation Daybreak.''}
\end{quote}
As explicitly noted in ETDA’s encyclopedia, these two Operation Daybreaks correspond to distinct campaigns and should not be merged. This warning was not considered in one of the mappings we use, causing these two name clusters to merge into one.

\subsection{Large clusters, weak associations, and cut edges}
\begin{figure}
\centering
\includegraphics[width=\linewidth]{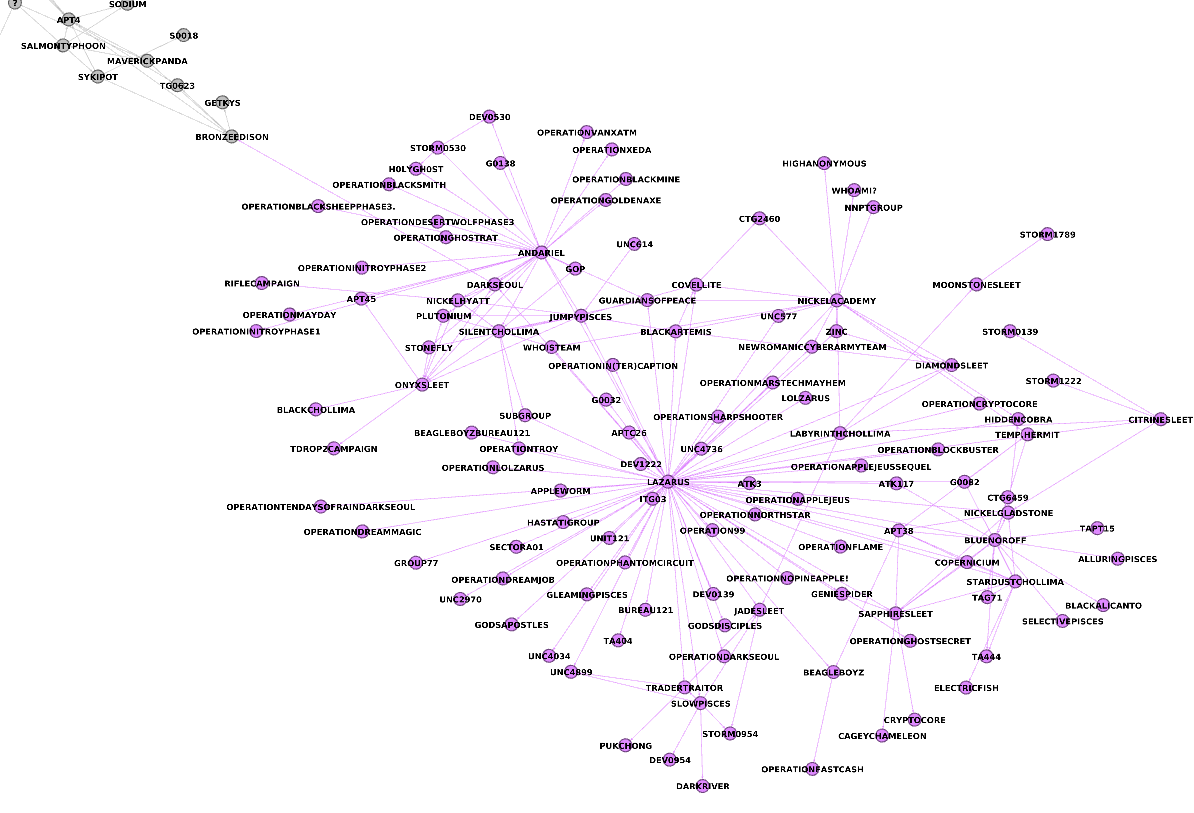}
\caption{Partial view of the largest cluster in the TANAG showing a cut edge between \texttt{UNK:OPERATIONDARKSEOUL} and \texttt{SECUREWORKS:BRONZEEDISON}.}
\label{fig:large_weak_cluster}
\end{figure}

The subgraph shown in Figure \ref{fig:large_weak_cluster} is part of the largest name cluster (\largestcluster) in the TANAG. This subgraph contains 520 nodes and includes well-known TA names such as \texttt{UNK:LAZARUS}, \texttt{UNK:ANDARIEL} and \texttt{PALOALTOUNIT42:JUMPYPISCES}, as well as multiple operation names typically linked to these TAs, such as \texttt{UNK:OPERATIONBLOCKBUSTER} or \texttt{UNK:OPERATIONDARKSEOUL}. The subgraph is connected to the 400 nodes that form the remainder of \largestcluster by a single connection: an edge between \texttt{UNK:OPERATIONDARKSEOUL} and \texttt{SECUREWORKS:BRONZEEDISON}. This connection can be traced to a SecureWorks post~\cite{secureworksBRONZEEDISON} profiling Bronze Edison, which states that Dark Seoul is an alias of Bronze Edison. 

The previous example constitutes a cut edge, or bridge, in the cluster, i.e., an edge that, if removed, the number of connected components increases. Cut edges represent critical points of vulnerability or single points of failure in a network. In the case of the TANAG as produced by \toolname, they are responsible for a substantial increase in the number of aliases attributed to the same TA. For example, if we consider the 514 nodes of the largest alias cluster and apply transitive closure (if \texttt{x} is an alias of \texttt{y} and \texttt{z} is an alias of \texttt{x}, then \texttt{z} is an alias of \texttt{y}), we obtain $131,841$ alias pairs. However, if we remove the edge between Dark Seoul and Bronze Edison discussed above, we obtain only $6,441$ alias pairs within the smaller subgraph and $79,800$ alias pairs in the resulting largest alias cluster, reducing the number of alias pairs by almost $35\%$ with a single edge removal.

Based solely on publicly available information, it is generally not possible to determine whether an association claimed by a vendor is correct or not. However, identifying cut edges in the graph provides a simple mechanism to list aliases responsible for the existence of large name clusters.

\subsection{Tools vs. Groups, or Winnti is not a TA}
\begin{figure}
\centering
\includegraphics[width=\columnwidth]{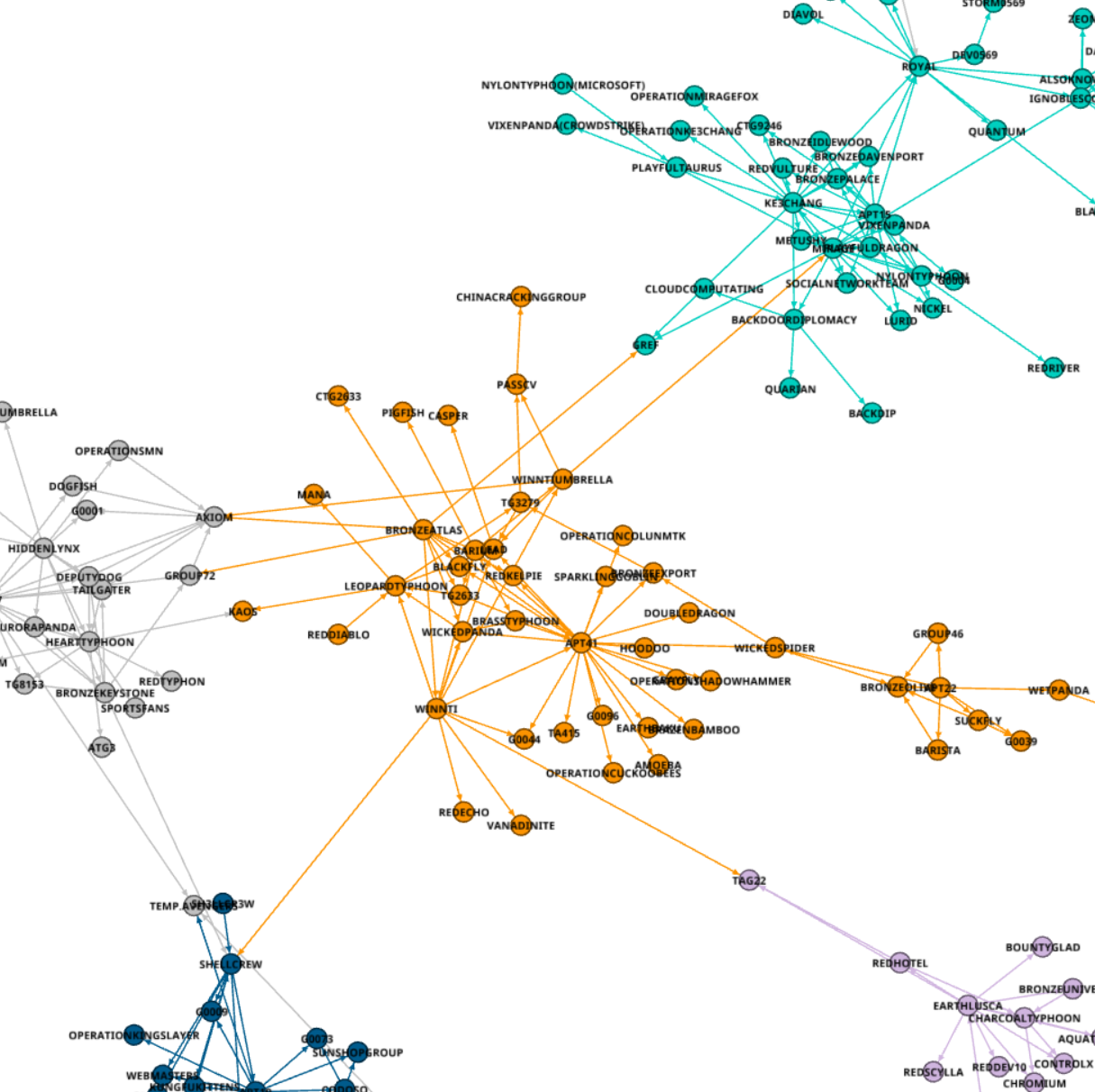}
\caption{Partial view of the alias cluster connecting multiple TA names with the \texttt{Winnti} designation.}
\label{fig:winnti}
\end{figure}

Figure~\ref{fig:winnti} shows another subgraph of \largestcluster~ comprising numerous interconnected nodes and edges. Unlike the previous case, no clear pattern emerges in the connectivity (i.e., edges do not segregate into well-defined communities). This is consistent with the fact that the nodes in the community displayed in orange are associated with both \texttt{UNK:WINNTI} and \texttt{UNK:WINNTIUMBRELLA}. As noted in Section~\ref{sec:introduction}, activity initially attributed to the Winnti group was later shown to involve multiple TAs reusing common tooling. Consequently, many edges in this subgraph capture associations between the legacy `Winnti' label and the specific TAs assessed to have conducted particular operations. One example is the edge connecting \texttt{UNK:WINNTIUMBRELLA} and \texttt{UNK:AXIOM}, as defined by the \texttt{APTGO} mapping. This association can be traced to a series of reports published by Novetta~\cite{WaybackNovettaSMNWinnti} in April 2015.

\subsection{The persistence of temporary names}
Some of the clusters in the TANAG include the temporary names that activity clusters were given upon discovery, typically linked to a designated TA name. For example, one cluster contains both \texttt{MANDIANT:UNC2452} and \texttt{MANDIANT:APT29}, while another contains \texttt{MANDIANT:UNC2630}, \texttt{MANDIANT:TEMP.BOTTLE} and \texttt{MANDIANT:APT5}. Labels using prefixes like \texttt{UNC} and \texttt{TEMP} are commonly used as a provisional name when analysts need to report on an activity cluster without providing a definitive attribution. These temporary designations serve as placeholders until further information clarifies the TA's identity. In cases where these activity clusters are subsequently revisited to link them to a designated TA, keeping the initial temporary names is often the only way of associating the original CTI product (report, IoCs, etc.) with the new one. However, the use of such temporary names significantly contributes to the overall naming mess surrounding threat actors, potentially adding layers of ambiguity and complicating efforts to achieve consistent attribution.

\section{Discussion: Threat Actors Are Not Stars}

``Celestial nomenclature has long been a controversial topic.'' This sentence opens the International Astronomical Union (IAU) guidelines on the naming of celestial objects~\cite{IAU_naming_celestial}. To get an idea of how far from being solved the naming problem in astronomy is, consider these two facts: planet names were not standardized until 1976, and the IAU Working Group on Star Names (WGSN), whose purpose is to formally catalog the names of stars, launched in 2016.

Researchers outside the CTI field may be tempted to think of threat actor activity as astronomical objects and seek similarities with the historical naming problem in astronomy. At the end of the day, both phenomena share a remote nature --one that we can only hypothesize about using signals obtained through imperfect instruments. However appealing parallelisms between space and cyberspace may be, the comparison is profoundly misleading, for they are very different media.

\subsection{The Nature of Signals in Cyberspace}

Astrometry --the study of measurements of the positions and movements of celestial objects-- provides the basis for reproducibility and independent verification about the position and properties of a celestial object. The first star catalogs developed by ancient Greek and Chinese astronomers contain coordinates for the position of each celestial object using different coordinate systems, some of them rather informal and inaccurate~\cite{IAU_naming_stars}. In contemporary history, star catalogs adopted various naming schemes based on alphanumeric designators that are used as official identifiers but all share a crucial key feature: the position and properties of the object, and not the proper name, are the central features of a catalog entry, because they enable locating the celestial object and finding the equivalent designator in a different catalog.

In contrast to astrometry, cyberspace telemetry is of a very different nature. CTI vendors collect and analyze attack signals to produce CTI products. These products contain knowledge about an (attributed or not) TA and its tradecraft, including intent, capabilities, and an activity profile typically focused on known attack campaigns and the tactics, techniques, and procedures (TTPs) it uses. As with any other intelligence product, access to different sets of data may result in a different characterization of the phenomenon. Multiple reasons contribute to the existing divergence in the definition and characterization of threat actors in cyberspace:
\begin{itemize}
\item First, threat actor activity is ephemeral: It occurs in a given region of the cyberspace-time continuum and is confined to it. A vendor without telemetry on the systems where the activity took place at the moment that it happened will not be able to independently verify other vendor's findings.
\item Second, telemetry can be incomplete and may miss crucial data. This can be the result of limitations in the scope of the collected signals, which may not include all relevant data to the analyst or may have limited visibility.
\item Third, telemetry and other data sources are private to each CTI vendor. Each vendor has its own unique visibility, and the overlap with the telemetry of other vendors may not be enough to reach similar conclusions. This is a formidable obstacle to agreeing in the first place about threat activity observations. Where one vendor sees activity attributed to one TA another vendor can see two different TAs or none. Further, there are financial and strategic incentives to keep threat intelligence private. Even though organizations do occasionally share threat intelligence, this generally takes place on a case-by-case basis. Further, in many countries such a sharing is forbidden by national security laws.
\end{itemize}

\subsection{A TA naming standard?}

Naming conventions may have different goals and properties. One typical goal of a good naming scheme is to provide unique names for different objects. Others aim at achieving uniqueness while providing rules for naming objects \emph{systematically}. Systematic names are often derived from an underlying namespace that classifies objects according to salient features or organizes them according to some structure. For example, some namespaces in computing and networking are trees in which an entity is assigned a node and is named after the unique path that goes from that node to the root. Hierarchical naming is convenient because it produces unique names, reflects the structure (or some salient attributes) of the object, and allows reuse of common subnames. Systematic naming schemes like these are very common in the sciences because, in addition to being unique and providing formation rules, they are typically scalable.

Based on these principles, MISP has recently proposed a naming standard for TAs~\cite{MISP24}. The document identifies the key issues that motivate the need for a universal scheme: lack of syntax rules, confusion between the TA name and its tools or operations, proliferation of aliases and informal mappings between them that lead to ambiguity, and so on. It concludes by proposing a set of guidelines for a common framework that could result in more consistency, interoperability, and transparency in the ecosystem.

While reaching consensus on a convention for naming TAs may be plausible, the key obstacle would be for industry to adopt it for the reasons described above --especially obstacles for sharing private intelligence.
In absence of a universal TA naming scheme, the CTI community will continue relying on more or less curated mappings between the various catalogs. For academics and practitioners in the field, as important as the correctness and completeness of those mappings is understanding that (i) a name is just a designation assigned to a number of activity clusters; (ii) these clusters could be caused by one or more TAs; and (iii) those TAs may have non-trivial relationships with the names of other TAs as reported by a different CTI vendor. This complexity complicates tasks such as creating integrated large-scale corpora of public CTI reports from different sources, which would facilitate measurement studies. Unfortunately, this problem does not seem to have an easy solution.

\subsection{Limitations and significance}

The analysis presented in this work is inherently limited by the completeness and precision of the TA names and mappings available in our knowledge base. \toolname is built on a knowledge base extracted from public sources, some of which are incomplete and may contain errors. Given the open-source nature of the tool, these limitations can be collectively identified and corrected, allowing the knowledge base to be collaboratively refined and updated with new vendor taxonomies and mappings. While adding more sources and correcting potential errors would result in a different set of alias clusters, we expect the overall conclusions discussed in this study to remain valid.

We believe that \toolname can serve as a valuable tool for numerous applications, but it must be used with caution, particularly in light of the issues detailed in Section \ref{sec:errors}. The resulting alias clusters  capture the outcome of integrating various mappings, but do not account for the presence of incorrect or questionable alias relationships within those mappings. Therefore, analysts must filter out or post-process these outputs according to their specific requirements.


\bibliographystyle{model1-num-names}

\bibliography{biblio}

\end{document}